\documentclass[onecolumn]{aastex631}

\usepackage{xcolor}
\usepackage{graphicx}
\usepackage{amsmath}
\usepackage{gensymb}
\usepackage{xcolor}

\begin{document}

\title{A Numerical Experiment on Oscillatory Magnetic Reconnection in a Laboratory Plasma System Driven by Alternating Currents}

\author{Sripan Mondal}
\affiliation{Department of Physics, Indian Institute of Technology (BHU), Varanasi-221005, India}
\author{Abhishekh Kumar Srivastava}
\affiliation{Department of Physics, Indian Institute of Technology (BHU), Varanasi-221005, India. Email:- asrivastava.app@itbhu.ac.in}
\author{Eric R. Priest}
\affiliation{Mathematics Institute, St Andrews University, KY16 9SS, St Andrews, UK}



\begin{abstract}
Using the open source MPI-AMRVAC framework, we study oscillatory reconnection in a laboratory plasma, which occurs when a magnetic null is perturbed by incoming fast magnetoacoustic waves driven by an alternating current. The magnetic null region collapses to first form a $y$-directed current sheet that later changes its orientation to the $x$-direction. The $x$-directed current sheet has smaller enhanced thermal pressure and out-of-plane current than the $y$-directed current sheet. The Hall effect produces an out-of-plane plasma flow that evolves with a time lag with respect to the enhanced thermal pressure and out-of-plane current density. Increasing the amplitude of the alternating current produces higher thermal pressure, out-of-plane current density, and out-of-plane plasma flow, while the first peaks of thermal pressure and out-of-plane current density occur earlier.
\end{abstract}

\keywords{Hall Magnetohydrodynamics; MHD Waves; Magnetic Reconnection; Laboratory Plasma}

\section{Introduction} 
Breaking and topological rearrangement of magnetic field lines, namely, magnetic reconnection, occurs in different plasma systems such as solar and space plasmas, planetary magnetospheres, and laboratory fusion plasmas \citep[e.g.,][]{Yamada2010,HesseCassak2020,Nakamura2025}. Such a process can result in the liberation of non-potential magnetic energy stored in the magnetic configurations and its transformation into heat, bulk kinetic energy of the plasma, and kinetic energy of charged particles \citep[e.g.,][]{Goedbloed2004,Priest2014}. In principle, magnetic reconnection is a three-dimensional process, which can take place near the magnetic nulls, separators, or quasi-separators configured in various kinds of magnetic geometry \citep[e.g.,][]{PontinPriest2022}. A two-dimensional description of magnetic reconnection requires a magnetic X-type geometry in which a magnetic null is located at the intersection of two separatrix curves \citep[e.g.,][]{PontinPriest2022}. However, the formation of a sheet of high current density (namely, a current sheet) is an essential step towards the onset of magnetic reconnection irrespective of the geometry.

Investigations of magnetic reconnection have been carried out using observational data from the space satellites for both solar \citep[e.g.,][]{Drake2025} and magnetospheric plasmas \citep[e.g.,][]{Gershman2024}. Several dedicated laboratory experiments such as MRX (Magnetic Reconnection Experiment) \citep[e.g.,][]{YamadaJi2005,Yamada2014,YamadaYoo2016} and TREX \citep[e.g.,][]{Olson2016} have been set up to improve our understanding of reconnection in the laboratory. There are also laser-induced high energy density experiments of magnetic reconnection \citep[e.g.,][]{Fox2011, Fiksel2014,Egedal2021,Fox2022}. In addition, there are experiments driven by pulsed-power platforms in which the plasma pressure is on the order of the magnetic pressure \citep[e.g.,][]{Hare2017}. 

In parallel, numerical simulations play a significant role in enhancing the existing understanding of magnetic reconnection at disparate spatio-temporal scales \citep[e.g.,][]{Pontin2024,Moreno2025,Shay2025,Richter2025}. Single-fluid magnetohydrodynamic (MHD) models are frequently used in two- or three-dimensional models. However, when ions and electrons are decoupled, a two-fluid prescription is preferable.  If the length-scale of plasma dynamics is larger than the electron inertial length but smaller than the ion inertial length, corrections due to the Hall effect need to be incorporated into the MHD equations, which is the focus of this paper \citep[e.g.,][]{Birn2001,huba02,huba03a,huba03b,Knoll2006,Malakit2009,Stanier2017}.

If the initiation is associated with MHD instabilities such as the resistive tearing mode or the ideal kink mode \citep[e.g.,][]{Baty2000,Browning2024}, it is termed spontaneous reconnection. However, if the reconnection is caused by some external driving such as flux emerging from below the solar surface, erupting prominences or magnetoacoustic waves  \citep[e.g.,][]{Sakai1984,Vekstein2017,Srivastava2019}, it is regarded as externally driven reconnection. In laboratory plasma systems,  most of the experiments have reconnection that is externally driven \citep[e.g.,][]{Yamada1997,Hare2017}.

Since various kinds of wave mode are ubiquitous in the solar corona, they can interact with nulls, separators or quasi-separators to initiate magnetic reconnection. This process, together with the possibility of reconnection driving waves, is referred to as a `Symbiosis of Waves and Reconnection (SWAR)' \citep[e.g.,][]{Mondal2024b,Srivastava2025}. At laboratory scales, such a process has been studied using an experiment in which an alternating current varying between 40 kA and 70 kA with a half-period of  400 $\mu$s generates fast magnetoacoustic waves that converge towards a magnetic null. This results in the accumulation of current to  form a current sheet that gradually thins in the presence of Hall currents and inverse currents \citep[e.g.,][]{Frank2024}.  

In the solar corona, if a current sheet changes its orientation from, say, vertical to horizontal or vice-versa, it is termed oscillatory reconnection \citep[e.g.,][]{Craig1991,McLaughlin2009,Thurgood2017,Karampelas2022}. However, the periodicity of such an oscillation depends on the background plasma parameters rather than the details of the driver and can take place in a self-consistent manner even in the absence of continuous driving \citep[e.g.,][]{Karampelas2023}. Moreover, in a reconnection experiment in MRX, it has been found that, during an increase of current in the poloidal flux loops from 0 to 15 kA, a vertical current sheet is formed, which undergoes `Push' reconnection. Then, during a decrease of the current, the sheet changes its orientation and undergoes `Pull' reconnection \citep[e.g.,][]{Yamada1997}.

Here, we conduct a two-dimensional Hall-MHD simulation of such externally-driven oscillatory reconnection in a plasma system that is magnetically dominated. Our main aim is to examine how physical variables change with time during the collapse of a magnetic null due an alternating current flowing through wires close to the boundary of the simulation domain. In section 2, the numerical setup is presented along with the adopted numerical methods. In section 3, the scientific outcomes are demonstrated and discussed in detail. This is followed by a summary of the findings in the section 4.

\section{Setup for the Numerical Experiment}
 
We consider a plasma system having a uniform number density ($n_{0}$) of $10^{13} ~\mathrm{cm}^{-3}$ and a temperature ($T_{0}$) of $10^{5}~\mathrm{K}$, i.e., around $8.6 ~\mathrm{eV}$, which are similar to those used in the MRX setup while conducting a study of collisional magnetic reconnection in a low-$\beta$ (i.e., magnetically dominated) plasma. The resulting Debye length ($\lambda_D = \sqrt{\epsilon_{0}k_{B}T_{0}/n_{0}e^{2}}$) is estimated to be around $7 \times 10^{-4}~\mathrm{cm}$, while the electron inertial length ($d_{e}=\sqrt{\epsilon_{0}m_{e}c^{2}/n_{0}e^{2}}$) and ion inertial length ($d_{i}=\sqrt{\epsilon_{0}m_{p}c^{2}/n_{0}e^{2}}$) are  0.17 cm and 7.2 cm, respectively. Here, $\epsilon_{0}$, $k_{B}$, $c$, $m_{e}$ and $m_{p}$ are the permittivity of free space, Boltzmann constant, speed of light, electron mass and proton mass, respectively. We consider a two-dimensional simulation domain spanning over $[-7.5d_{i}, 7.5d_{i}]$, (i.e., $[-54~\mathrm{cm}, 54~\mathrm{cm}]$) in both the $x$- and $y$-directions. 

\begin{figure*}
    \centering
    \includegraphics[width=1\linewidth]{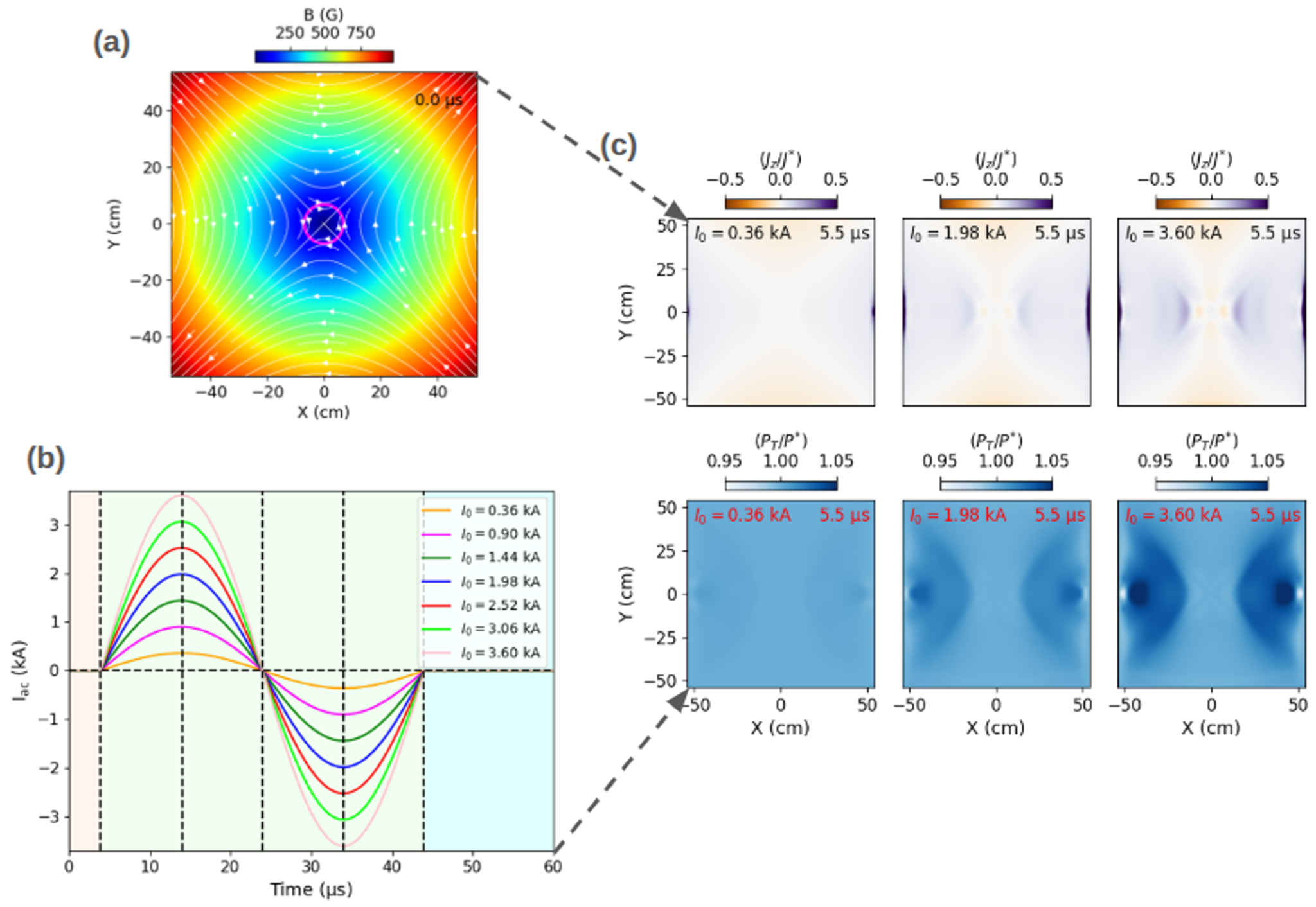}
    \caption{(a) The initial magnetic field configuration, in which the $\beta$ = 1 curve where the thermal ($P_{T}$) and magnetic ($P_{M}$) pressure are equal is denoted by a magenta circle. (b) The variation with time of the alternating currents of different amplitude imposed at the left and right side boundaries; vertical  dashed lines correspond to times of 3.9 $\mu$s, 13.9 $\mu$s, 24 $\mu$s, 34 $\mu$s and 44 $\mu$s. (c) The perturbations in magnetic field and plasma as wave-like fronts seen in the $z$-component of current density ($J_{z}$) (top row) and $P_{T}$ (bottom row) at 5.5 $\mu$s for three different amplitudes of the alternating current. The inward propagation of the initial wave-like perturbations from  0 to 7.6 $\mu$s is shown in Fig1c.mp4 (Multimedia available online).}
    \label{fig:fig1}
\end{figure*}

We place four straight wires each having length ($L$) of 360 cm,  extending from $z$ = $-180$ cm to $z$ = +180 cm in the $z$-direction at the following locations: $(x_{1}, y_{1})$ = $[0~\mathrm{cm}, -216~\mathrm{cm}]$, $(x_{2}, y_{2})$ = $[0~\mathrm{cm}, +216~\mathrm{cm}]$, $(x_{3}, y_{3})$ = $[-216~\mathrm{cm}, 0~\mathrm{cm}]$ and $(x_{4}, y_{4})$ = $[+216~\mathrm{cm}, 0~\mathrm{cm}]$. Direct currents of amplitude $I_{dc}$ = 72 kA flow through each wire in the positive $z$-direction. The resulting magnetic field possesses a magnetic null patch at the middle of the simulation domain in the $z$ = 0 plane (see Fig.~\ref{fig:fig1}(a)). If we define
\begin{equation}
r_{i}^{2} = (x-x_{i})^{2}+(y-y_{i})^{2}
\label{eq1}
\end{equation}
for $i =1, 2, 3, 4, 5, 6$, then the $x$-component of the magnetic field in the $z$ = 0 plane can be written as--
\begin{equation}
B_{0x} =  -\sum_{i=1}^{2} \frac{I_{dc}L(y-y_{i})}{r_{i}^2 \times \sqrt{r_{i}^{2}+L^{2}/4}} + \sum_{i=3}^{4} \frac{I_{dc}L(y-y_{i})}{r_{i}^2 \times \sqrt{r_{i}^{2}+L^{2}/4}},
\label{eq2}
\end{equation}
while the $y$-component in the same plane is
\begin{equation}
B_{0y} =  \sum_{i=1}^{2} \frac{I_{dc}L(x-x_{i})}{r_{i}^2 \times \sqrt{r_{i}^{2}+L^{2}/4}} - \sum_{i=3}^{4} \frac{I_{dc}L(x-x_{i})}{r_{i}^2 \times \sqrt{r_{i}^{2}+L^{2}/4}}.
\label{eq3}
\end{equation}
Since, these wires are far away from the boundaries of the simulation domain, the resulting magnetic field is in quasi-static equilibrium in the presence of a uniform density and temperature, if it is not subjected to any other perturbations.

We place two more straight $z$-directed wires of length  $L_{1}$ = 3.6 cm, along which alternating currents flow at the following locations: $(x_{5}, y_{5})$ = $[-57.6~\mathrm{cm}, 0~\mathrm{cm}]$, and $(x_{6}, y_{6})$ = $[57.6~\mathrm{cm}, 0~\mathrm{cm}]$. The periodicity and phase of the alternating currents flowing through them are the same (see Fig.~\ref{fig:fig1}(b)) and have the form
\begin{equation}
I_{ac} = I_{0}f(t),
\label{eq4}  
\end{equation}
\quad \textrm{where} \quad
\begin{equation}
f(t) =
\begin{cases}
0 & \text{when $t \leq t_1$;} \\
\sin\left(\frac{2\pi(t-t_{1})}{T}\right) & \text{when $t_1 \leq t \leq t_2$;} \\
0 & \text{when $t> t_2$,}
\end{cases}
\label{eq5}
\end{equation}
where $t_1 = 4$~$\mu$s, $t_2 = 44$~$\mu$s and $T = 40$~$\mu$s. We vary the amplitude $I_{0}$ from 0.36 kA to 3.60 kA to see how it affects the dynamics of the reconnection. The alternating currents generate a time-varying magnetic field given by
\begin{equation}
b_{x} = \frac{I_{ac}L_{1}(y-y_{5})}{r_{5}^2 \times \sqrt{r_{5}^{2}+L_{1}^{2}/4}}+\frac{I_{ac}L_{1}(y-y_{6})}{r_{6}^2 \times \sqrt{r_{6}^{2}+L_{1}^{2}/4}},
\label{eq6}
\end{equation}
\begin{equation}
b_{y} = -\frac{I_{ac}L_{1}(x-x_{5})}{r_{5}^2 \times \sqrt{r_{5}^{2}+L_{1}^{2}/4}}-\frac{I_{ac}L_{1}(x-x_{6})}{r_{6}^2 \times \sqrt{r_{6}^{2}+L_{1}^{2}/4}}.
\label{eq7}
\end{equation}

We discretize the simulation domain using a resolution of $160 \times 160$ initially, which is then subjected to one level of adaptive mesh refinement to have an effective maximum resolution of $320 \times 320$ with the smallest cell size being 0.34 cm. Since, this smallest cell size is much smaller than the ion inertial length of 7.2 cm, the Hall effect decouples the ions and electrons and is well treated in our simulation. The smallest cell size is nearly twice the electron inertial length of 0.17 cm, but a full kinetic simulation would treat the electron behaviour better. Thus, we solve the following dimensionless Hall-MHD equations to simulate the magneto-plasma dynamics using open-source MPI-AMRVAC \citep[e.g.,][]{Keppens2023}: 
 \begin{equation} 
\frac{\partial \rho}{\partial t} + \vec{\nabla} \cdot ( \rho \vec{V} ) = 0,
\label{eq8}
\end{equation}
\begin{equation}
  \frac{\partial}{\partial t}(\rho \vec{V}) + \vec{\nabla} \cdot \left [ \rho \vec{V}\vec{V}  + P_{tot}\vec{I} - \vec{B}\vec{B} \right ] = 0 ,
\label{eq9}
\end{equation}

\begin{equation}
\begin{split}
\frac{\partial e}{\partial t} +  \vec{\nabla} \cdot \left( e\vec{V} + P_{tot}\vec{V} -\vec{B}\vec{B} \cdot \vec{V}\right)+\\
\vec{\nabla} \cdot \left(\frac{\eta_{H}}{\rho}[(\vec{J} \cdot \vec{B})\vec{B}-(\vec{B}.\vec{B})\vec{J}]\right)= 0,
\end{split}  
\label{eq10} 
\end{equation}

\begin{equation}
\frac{\partial \vec{B}}{\partial t} + \vec{\nabla} \cdot \left(\vec{V}\vec{B} - \vec{B}\vec{V}+ \frac{\eta_{H}}{\rho} (\vec{B}\vec{J}-\vec{J}\vec{B})\right) = 0,
\label{eq11}
\end{equation}

\quad \textrm{where} \quad
\begin{equation}
P_{tot} = P_{T} + \frac{B^2}{2}, ~~e = \frac{P_{T}}{\gamma-1} + {\textstyle{\frac{1}{2}}\rho V^{2}} + \frac{B^2}{2}
\label{eq12}
\end{equation}
\quad \textrm{and} \quad
\begin{equation}
  \vec{J} = \vec{\nabla} \times \vec{B}, ~~\vec{\nabla} \cdot \vec{B} =0.
  \label{eq13}
\end{equation} 
Here, $\eta_{H}$ = $\rho/(n_{e}e)$, and the dimensionless value of $\eta_{H}$ is derived from a normalized background density \citep[e.g.,][]{Leroy2017}. So, since the normalized background density in the present case is 1,   the (dimensionless) value of $\eta_{H}$ to be 1. We do not include a physical resistivity in the simulation.

Since Eqs. (\ref{eq8})-(\ref{eq13}) are solved numerically, the variables have been made dimensionless by dividing the dimensional values by characteristic values. For example, the normalising length $L^{*}$ is here the ion inertial length ($7.2~\mathrm{cm}$). The corresponding temperature $T^{*}$ and number density $n^{*}$ are taken to be $10^{5}~\mathrm{K}$ and $10^{13}~\mathrm{cm^{-3}}$, respectively, whereas the  density $\rho^{*}=1.4n^{*}m_{H}$ is  $2.34\times10^{-11}~\mathrm{g~cm^{-3}}$, in terms of the hydrogen mass ($m_{H}$). The thermal pressure $P^{*} = 2.3n^{*}k_{B}T^{*}$ is $317.5~\mathrm{dyne~cm^{-2}}$, while the  other variables are magnetic field $B^{*}$ = $\sqrt{4\pi P^{*}}$ = $63~\mathrm{G}$,  velocity $V^{*} = B^{*}/(\sqrt{4\pi \rho^{*}})$ = $3.68\times10^{6}~\mathrm{cm~s^{-1}}$ and  time $t^{*}=L^{*}/V^{*}$ = 2 $\mu$s. 

Temporal integration is carried out using a $``$two-step$"$ method, and the $``$Harten-Lax-van Leer (HLL)$"$ Riemann solver \citep[e.g.,][]{Harten1983} is utilized to estimate the flux at cell interfaces. A second-order symmetric total variation diminishing limiter, namely $``$vanleer$"$ \citep[e.g.,][]{vanleer1979} is used to suppress spurious numerical oscillations. Thermal pressure and density are fixed at their  initial values at all four boundaries. On the side boundaries, the $x$-component of velocity is assumed to be antisymmetric, while the other two components are taken to be symmetric. On the top and bottom boundaries, the $y$-component of velocity is assumed to be antisymmetric with the other two components being symmetric. This ensures that the normal velocity vanishes at all four boundaries, which therefore reflect the plasma flows. Furthermore, we do not introduce any damping layers to damp the reflections. Hence, the boundary effects  play a role in the dynamics under consideration, especially, in the absence of external driving.  The imposed boundary conditions for the other variables except the magnetic field are similar to those considered previously \citep[e.g.,][]{Karampelas2022}. 
%
\begin{figure*}
    \centering
    \includegraphics[width=1\linewidth]{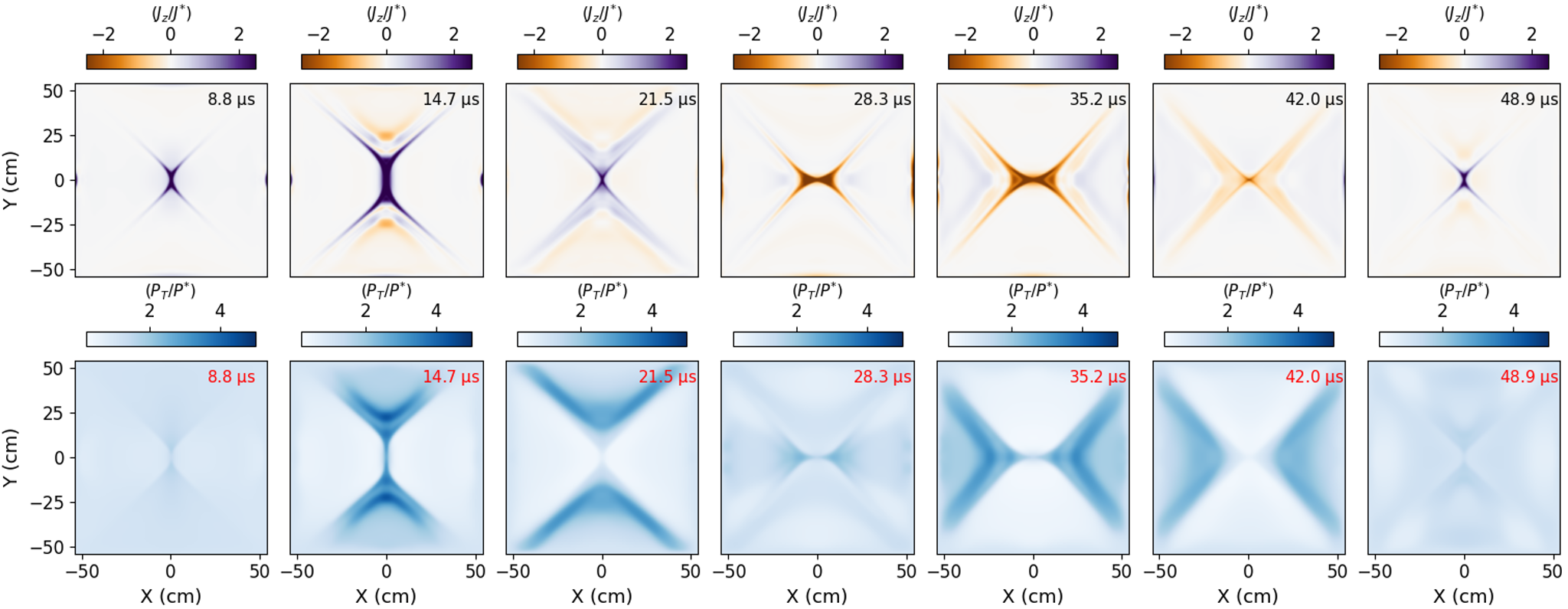}
    \caption{Variation with time of $J_{z}$ (top row) and $P_{T}$ (bottom row) for an imposed alternating current having an amplitude of 1.98 kA. $J_{z}$ changes both its magnitude and direction, while the current sheet changes its orientation and back again. The final column shows the continuation of the oscillation after the driver has been switched off. The changes in $J_{z}$ and $P_{T}$ from 7.8 to 60 $\mu$s are shown in Fig2.mp4 (Multimedia available online).}
    \label{fig:fig2}
\end{figure*}
%
\begin{figure*}
    \centering
    \includegraphics[width=0.95\linewidth]{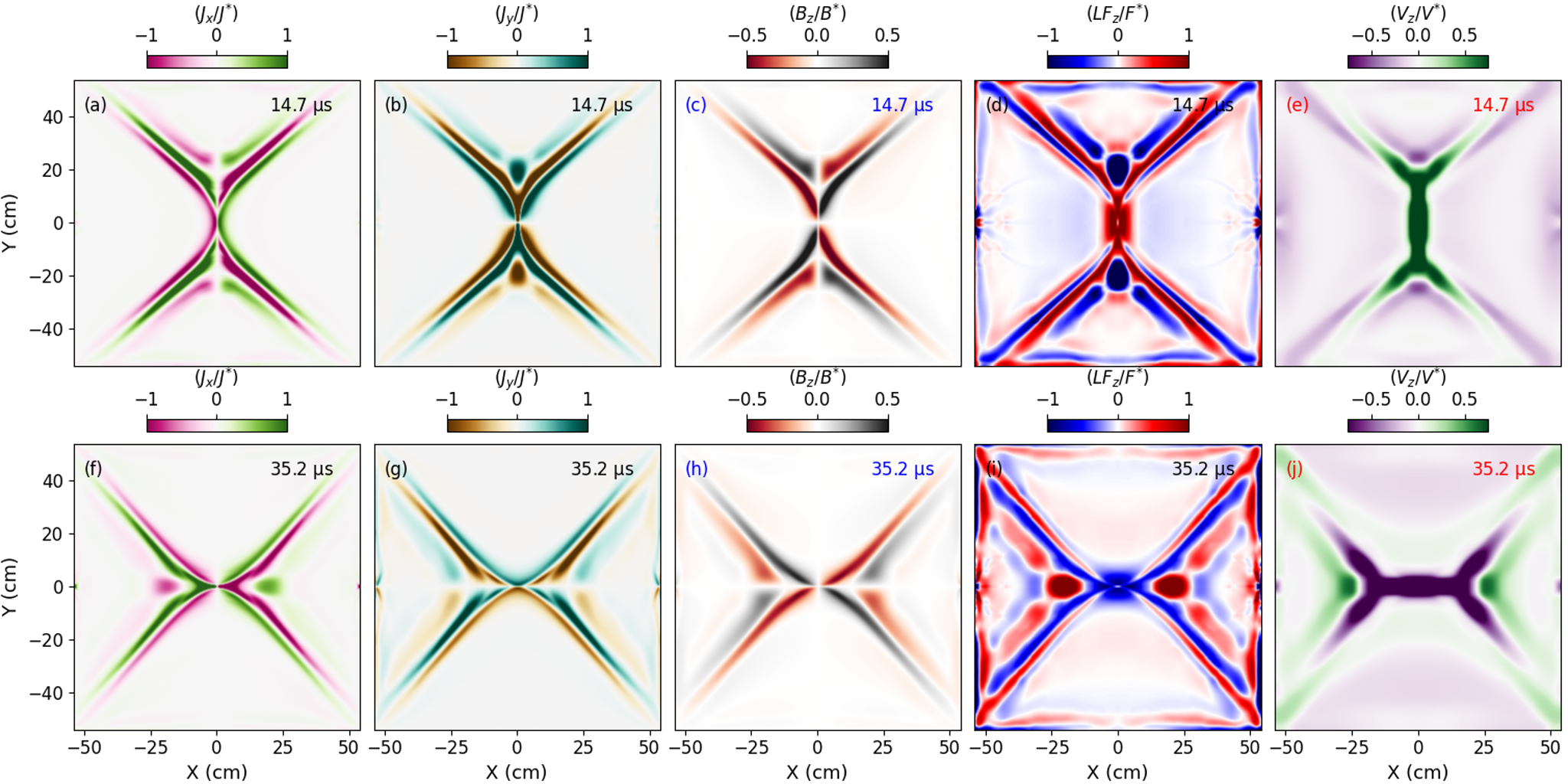}
    \caption{Spatial distribution of the in-plane current density components ($J_{x}$ (first column) and $J_{y}$ (second column)), out-of-plane magnetic field ($B_{z}$) (third column), out-of-plane Lorentz force ($LF_{z}$) (fourth column), and out-of-plane plasma flow ($V_{z}$) (fifth column) at 14.7 (top row) and 35.2 $\mu$s (bottom row), when the imposed alternating current amplitude is 1.98 kA. The evolutions of $J_{x}$, $J_{y}$, $B_{z}$, $LF_{z}$, and $V_{z}$ from 7.8 to 60 $\mu$s are shown in Fig3.mp4 (Multimedia available online).}
    \label{fig:fig3}
\end{figure*}

\section{Results}

Since the wires carrying the alternating currents are placed very close to the left and right side boundaries, such currents generate magnetic field perturbations, which then propagate inward into the simulation domain as fast magnetoacoustic waves and disturb the quasi-stable equilibrium. The waves can be modeled as perturbations in the background magnetic field denoted via out-of-plane current density (say, $J_{z}$) and in the background plasma conditions such as in thermal pressure ($P_{T}$) (see Fig.~\ref{fig:fig1}(c) (Multimedia available online)). The waves are found to reach the magnetic null around 6.5 to 6.8 $\mu$s after starting their propagation at 4 $\mu$s from the alternating current carrying wires giving a propagation speed of 207 to 230 $\mathrm{km\  s^{-1}}$. The amplitude of the perturbation increases with the amplitude of the driving alternating current (see Fig.~\ref{fig:fig1}(c)). Approaching fast magnetoacoustic waves cause the magnetic null point to collapse and form a current concentration at the null location and along the separatrices. In the following subsection, we discuss the evolution in $J_{z}$ and $P_{T}$ with an alternating current of amplitude $I_{0}$ = 1.98 kA, determine the role of Hall effects on the dynamics, and further demonstrate the oscillation in the out-of-plane electric field.

\subsection{Morphological Overview and Electric Field Oscillations for a Particular $I_{0}$}

Due to the resulting unbalanced Lorentz force, the magnetic null eventually forms a $y$-directed current sheet with a strong current ($J_{z}$) directed out of the plane. The magnitude of $J_{z}$ at first increases before beginning to decrease around 21 $\mu$s (see Fig.~\ref{fig:fig2} (Multimedia available online)). From 24 $\mu$s to 44 $\mu$s, the orientation of the current sheet changes  to horizontal (i.e., the $x$-direction) along with a reversal of the direction of $J_{z}$. Such a change in the orientation of the current sheet is a manifestation of oscillatory reconnection. Comparatively weaker inverse $J_{z}$ profiles are evident in the outflow regions at 14.7 $\mu$s and 35.2 $\mu$s, associated with strong reconnection outflows. Moreover, even though the driving alternating current is switched off around 44 $\mu$s, the cycle of current accumulation and reversal continues for some time (see right-most column of top row of Fig.~\ref{fig:fig2}). Such a persistence of oscillatory reconnection has previously been discussed \citep[e.g.,][]{Craig1991,McLaughlin2009,Karampelas2022,Karampelas2023} and can occur also in the absence of a Hall term. 

As discussed above, changes in $J_{z}$ with time can give us an indication of how the magnetic topology changes in time. $P_{T}$ increases within the current sheet structure and also along the separatrices with increasing $J_{z}$ (see Fig.~\ref{fig:fig2}), as well as in the outflow regions. $P_{T}$ decreases within the current sheet and along the separatrices when $J_{z}$ decreases. However, a higher value of $P_{T}$ persists in the outflow regions at farthest distances from magnetic Y-points and separatrices (see third and sixth column of bottom row of Fig.~\ref{fig:fig2}) most likely due to the combined effect of resistive heating and presence of Hall term. At later stages, after 44 $\mu$s, $P_{T}$ decreases almost to its initial value, which is similar to purely resistive behaviour.

To understand the role of the Hall effect, we consider the spatial distributions of in-plane currents ($J_{x}$ and $J_{y}$) and out-of-plane magnetic field ($B_{z}$) (see Fig.~\ref{fig:fig3} (Multimedia available online)), which vanish in resistive MHD but are a common feature of Hall-current models \citep[e.g.,][]{huba02,huba03a}. Unlike $J_{z}$, the in-plane currents exist not within the current sheet but at its outer peripheries. In addition to the accumulation of in-plane currents along the separatrices, inverse $J_{x}$ and $J_{y}$ are present in the outflow region at 14.7 $\mu$s and 35.2 $\mu$s, respectively (see Figs.~\ref{fig:fig3}(a) and (g)). Moreover, when a $y$-directed current sheet is present, there are localized peaks in $|J_{y}|$ along the outflow regions at 14.7 $\mu$s (see green and orange dotted structures in Fig.~\ref{fig:fig3}(b)). Similarly, for an $x$-directed current sheet, there are localized peaks in $|J_{x}|$ along the $y$ = 0 line in the outflow regions at 35.2 $\mu$s (see green and magenta dotted structures in Fig.~\ref{fig:fig3}(f)). These additional features are associated with the presence of inverse $J_{z}$ in the outflow regions \citep{Frank2024}.

The Hall current profile produces a quadrupolar configuration for $B_{z}$ near the diffusion region, typical of Hall effects \citep[e.g.,][]{huba02,huba03a,huba03b,Shay2001,Uzdensky2006,Drake2008}. Moreover, at times 14.7 $\mu$s and 35.2 $\mu$s, a more complex nested $B_{z}$ is found. Apart from the quadrupolar field concentrated along the magnetic separatrices, another set of oppositely oriented quadrupolar magnetic fields appears in the outflow regions in association with the inverse Hall currents (see Figs.~\ref{fig:fig3}(c), (h)). Therefore, the variation of this inverse $B_{z}$ with $x$ in the outflow regions near the peaks in $|J_{y}|$ possesses a high gradient as it reverses direction over a short distance across the peaks in $|J_{y}|$ (see Figs.~\ref{fig:fig3}(b)-(c)). Similarly, across the peaks in $|J_{x}|$, there are high gradients of inverse $B_{z}$ along the $y$-direction (see Figs.~\ref{fig:fig3}(f), (h)).

In pure 2D MHD simulations in the $xy$-plane without any guide field, the Lorentz force will only have $x$- and $y$-components. However, the presence of in-plane Hall currents produces a non-vanishing Lorentz force in the $z$-direction (say, $LF_{z}$). Its direction  within the current sheet changes from positive to negative when the sheet alters its orientation from the $y$- to the $x$-direction (see Figs.~\ref{fig:fig3}(d), (i)). The out-of-plane plasma flow ($V_{z}$) inside the current sheet also changes its direction  (see Figs.~\ref{fig:fig3}(e), (j)). The directional similarities between $LF_{z}$ and $V_{z}$ are also present along the separatrices (see Fig.~\ref{fig:fig3}). Moreover, there are complex configurations of in-plane currents in the outflow regions which produce weaker inverse $LF_{z}$ and $V_{z}$ there (see Fig.~\ref{fig:fig3}). 

\begin{figure}
    \centering
    \includegraphics[width=0.65\linewidth]{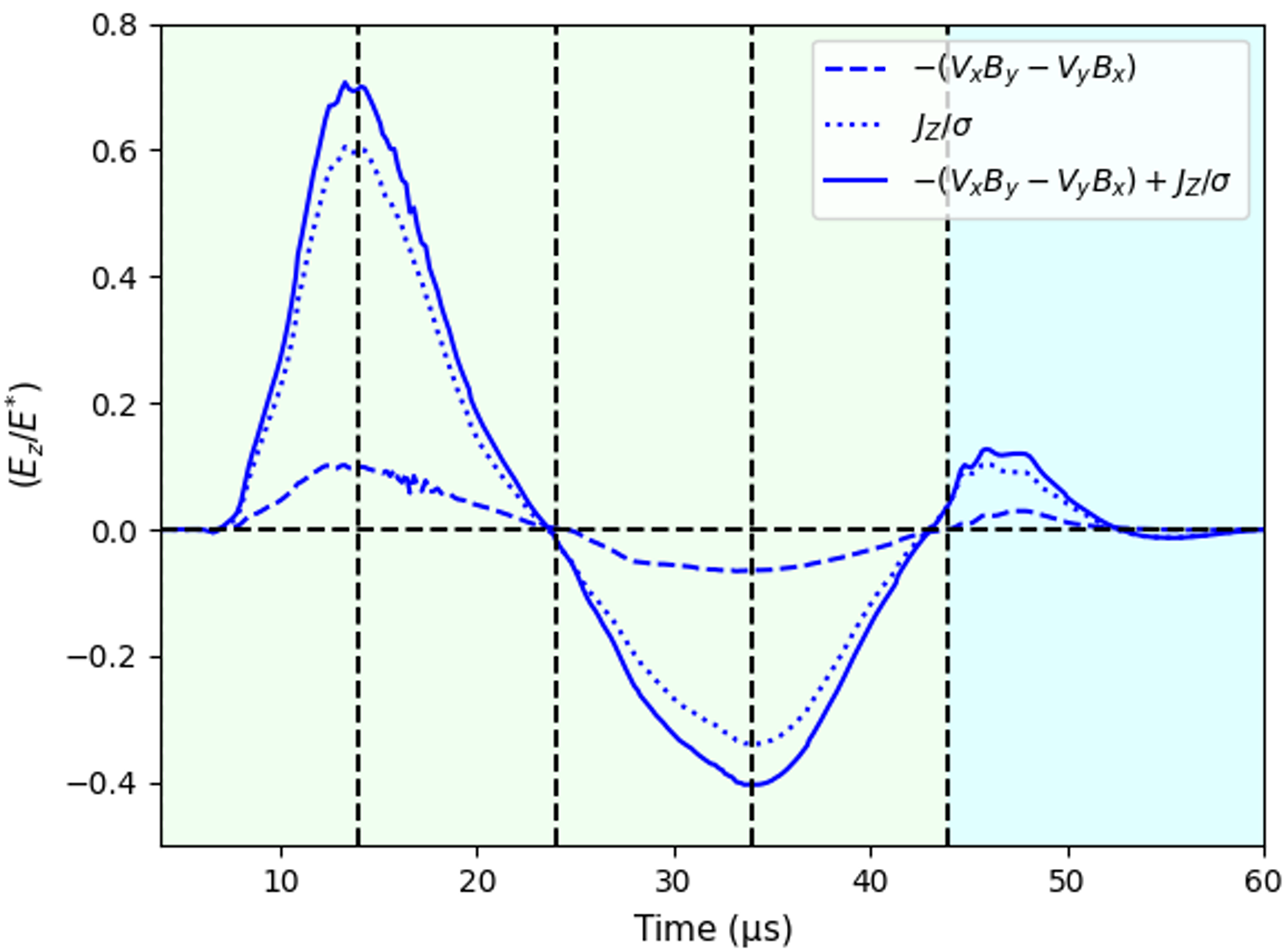}
    \caption{Oscillations in the out-of-plane electric field, i.e., $E_{z}$ at ($x,y$) = [0,0] cm, i.e., at the primary location of the magnetic null for $I_{0}$ = 1.98 kA. The conductive, diffusive and total $E_{z}$ are shown as dashed, dotted and solid blue curves respectively. Vertical black dashed lines denote 13.9 $\mu$s, 24 $\mu$s, 34 $\mu$s and 44 $\mu$s, as in Fig.~\ref{fig:fig1}(b).}
    \label{fig:fig4}
\end{figure}

To further ensure occurrence of the oscillatory reconnection, we examine whether the out-of-plane electric field, i.e., $E_{z}$ possesses any oscillation or not at the primary location of the magnetic null, i.e., at ($x,y$) = [0,0] cm while keeping $I_{0}$ fixed at 1.98 kA. Basically, we calculate $E_{z} = -(V_{x}B_{y}-V_{y}B_{x})+J_{z}/\sigma$. Here, $V_{x}$ and $V_{y}$ are the $x$- and $y$-components of the plasma velocity. Similarly, $B_{x}$ and $B_{y}$ are the $x$- and $y$-components of the magnetic field. $\sigma$ is the conductivity. We find that the diffusive part of $E_{z}$, i.e., $J_{z}/\sigma$ is much higher than the conductive part at the location of measurement (see dashed and dotted curves in Fig.~\ref{fig:fig4}). More importantly, we find that both diffusive and conductive parts of $E_{z}$ oscillate and change their direction with a reversal of the direction of the driving current (see Fig.~\ref{fig:fig4}). Therefore, we confirm that oscillatory reconnection is taking place at the location of magnetic null. Moreover, this oscillation in $E_{z}$ is indicative of the variation in energy conversion rate with time.

\begin{figure*}
    \centering
    \includegraphics[width=1\linewidth]{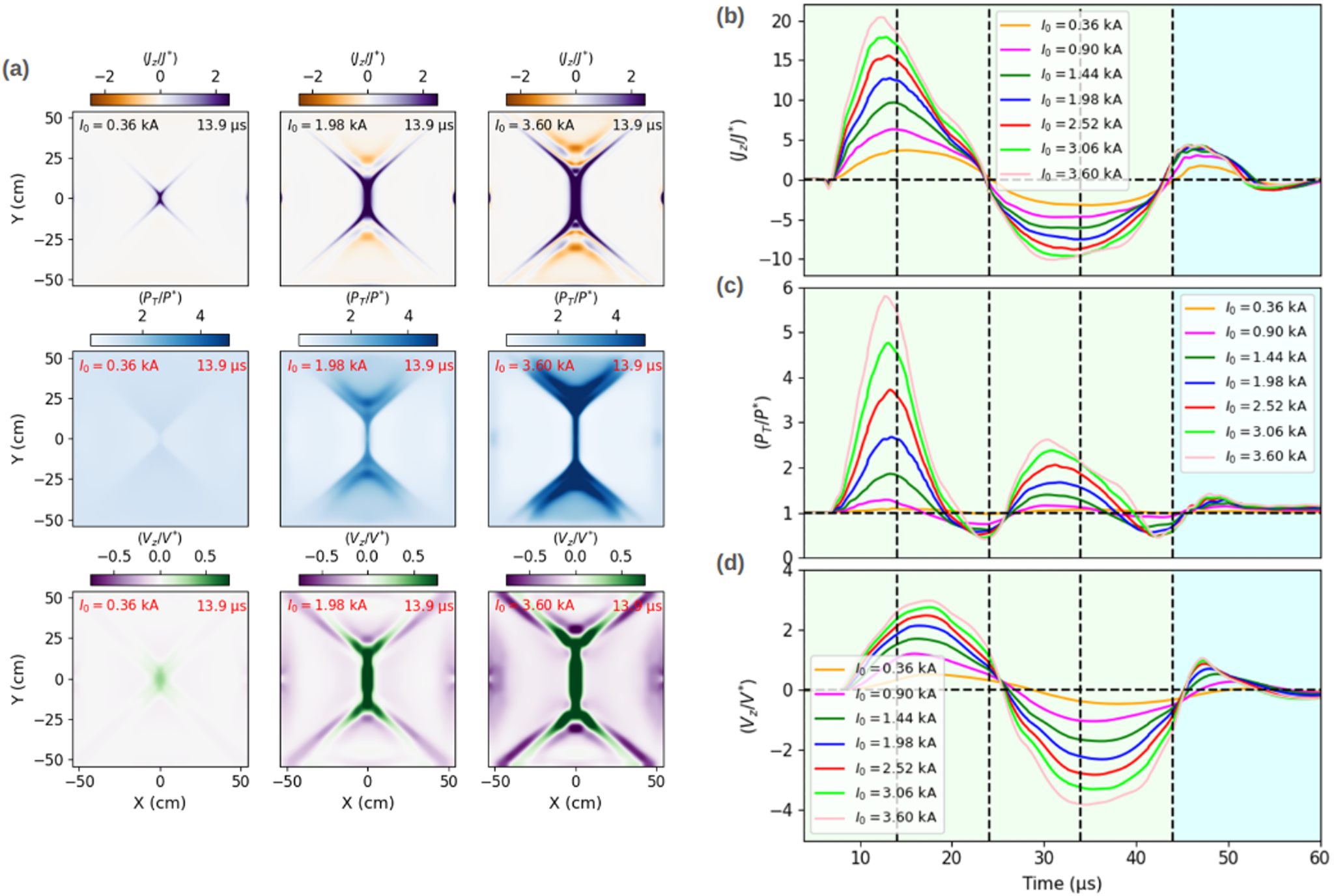}
    \caption{(a) $J_{z}$, $P_{T}$, and $V_{z}$ maps at 13.9 $\mu$s for imposed alternating currents of three different amplitudes (0.36 kA, 1.98 kA and 3.60 kA) (top to bottom). The evolutions of $J_{z}$, $P_{T}$, and $V_{z}$ from 7.8 to 60 $\mu$s for all three amplitudes are shown in Fig5a.mp4 (Multimedia available online). (b) The time-variation in  $J_{z}$ at the magnetic null ($x,y$) = [0,0] cm. Vertical black dashed lines correspond to 13.9 $\mu$s, 24 $\mu$s, 34 $\mu$s and 44 $\mu$s, as in Fig.~\ref{fig:fig1}(b) and in Fig.~\ref{fig:fig4}. (c) Variation in $P_{T}$ with time at the same location. (d) Variation in $V_{z}$ at the same location with time. Vertical black dashed lines in (c) and (d) are the same as in (b).}
    \label{fig:fig5}
\end{figure*}

\subsection{Temporal Evolution of $J_{z}$, $P_{T}$ and $V_{z}$ at the Primary Null Location and its Dependence on $I_{0}$}

Increasing the amplitude $I_{0}$ of the imposed currents produces a stronger perturbation in the magnetic field and results in more intense fast-mode waves (see Fig.~\ref{fig:fig1}(c)) that perturb the magnetic null to form a current sheet. The oscillatory reconnection remains similar, with a current sheet initially forming in the $y$-direction followed by one in the $x$-direction. We examine how $J_{z}$, $P_{T}$, and $V_{z}$ change within the current sheet and its surroundings if we change $I_{0}$ (see Fig.~\ref{fig:fig5}(a) (Multimedia available online)).  The magnitude of $J_{z}$ within the current sheet increases with an increase in $I_{0}$ (see the top row of Fig.~\ref{fig:fig5}(a)). $J_{z}$ within the current sheet in the absence of driving currents follows the same trend (see the animation associated with Fig.~\ref{fig:fig5}(a)). To have a more quantitative comparison, we calculate the change in $J_{z}$ with time at the origin ($x,y$) = [0,0] cm between 3.9 $\mu$s and 60 $\mu$s. Irrespective of $I_{0}$, $J_{z}$ begins to accumulate around 7.6 $\mu$s at the null location, and, after that, it keeps on increasing. The rate of increase of $J_{z}$ is found to be higher for higher $I_{0}$ (see Fig.~\ref{fig:fig5}(b)). For comparatively lower values of $I_{0}$, $J_{z}$ attains a peak around 13.9 $\mu$s, i.e., the time when sinusoidal driving currents reach their peak (denoted by the first vertical black dashed line in Fig.~\ref{fig:fig5}(b)). However, as $I_{0}$ is increased, the peaks in $J_{z}$ are attained at earlier times (see Fig.~\ref{fig:fig5}(b)). This may indicate that the dynamics are being driven by forces generated within the simulation domain rather than just responding directly to the boundary forcing.

In all of the cases, $J_{z}$ reaches zero around 24 $\mu$s and thereafter reverses its direction irrespective of the magnitude of $I_{0}$. This time of reversal for $J_{z}$ is similar to the time of reversal of the driving alternating current (denoted by the second vertical black dashed line in Fig.~\ref{fig:fig5}(b)). In comparison to the first phase with a $y$-directed current sheet, the magnitude of $J_{z}$ increases with a shallower slope for all $I_{0}$ in the second phase with the $x$-directed current sheet and attains lower magnitudes (see Fig.~\ref{fig:fig5}(b)). However, slopes of change in $J_{z}$ are definitely higher in the case of higher $I_{0}$ in both phases (see Fig.~\ref{fig:fig5}(b)). We do not even find proper dips in $J_{z}$ for any of the $I_{0}$ around 34 $\mu$s, i.e., the time when the driving current attains its dip (denoted by the third vertical black dashed line in Fig.~\ref{fig:fig5}(b)). Eventually, the magnitude of $J_{z}$ again decreases and reaches zero at 44 $\mu$s for lower $I_{0}$ (denoted by the fourth vertical black dashed line in Fig.~\ref{fig:fig5}(b)). Interestingly, $J_{z}$ reverses its direction even before 44 $\mu$s for higher $I_{0}$. The persistence of nonzero values of $J_{z}$ after the driving current is switched off is evident from the profiles of $J_{z}$ in the cyan-shaded region of Fig.~\ref{fig:fig5}(b). $J_{z}$ undergoes another oscillation in direction before declining to zero around 60 $\mu$s.  After the external driving is switched off, the profiles of $J_{z}$ for $I_0$ of 0.36 kA and 0.90 kA have lower magnitudes than for higher values of $I_{0}$. As discussed before, the contribution of the diffusive part, i.e., $J_{z}/\sigma$ is higher to the total $E_{z}$. Therefore, the dependence of the temporal variation of $J_{z}$ on $I_{0}$ is indicative of how the value of $I_{0}$ can change the features of  oscillatory $E_{z}$.

Next, we examine how the magnitude of $I_{0}$ affects the temporal variation of $P_{T}$ at ($x$,$y$) = [0,0] cm. From the colormaps of $P_{T}$ in Fig.~\ref{fig:fig5}(a), it is evident that an increase in $I_{0}$ increases $P_{T}$ within the current sheet, along the separatrices and also in the outflow regions. At ($x$,$y$) = [0,0] cm, $P_{T}$ starts to increase almost around 7.6 $\mu$s irrespective of $I_{0}$, similar to $J_{z}$. As for $J_{z}$ in Fig.~\ref{fig:fig5}(b), the slope of increase of $P_{T}$ is higher for higher $I_{0}$. Moreover, the first peaks in $P_{T}$ are also attained at earlier times than in 13.9 $\mu$s for higher $I_{0}$ (see Fig.~\ref{fig:fig5}(c)). After the peaks, $P_{T}$ at ($x$,$y$) = [0,0] cm falls to its initial background value even before the driving current vanishes during its reversal (see Fig.~\ref{fig:fig5}(c)). At times close to the reversal of the driving current, $P_{T}$ decreases to values even smaller than its initial value. Afterwards, during the increase of the driving current in the opposite direction, $P_{T}$ increases again up to roughly three times its initial value for $I_0$ = 3.60 kA. In general, the peaks in this second phase of reconnection are much smaller than those in the first phase. Similar to the first phase, $P_{T}$ decreases to its initial background value before the driving current is switched off at 44 $\mu$s for all $I_{0}$. In fact, as time advances, the value of $P_{T}$ first decreases beyond its initial value and then increases to reach near its initial value at times close to 44 $\mu$s. Later, it increases again beyond its initial value before eventually reaching an equilibrium around 60 $\mu$s. In general, an increase in amplitude of the driving current, $I_{0}$, raises the magnitude of $P_{T}$ at each time (see Fig.~\ref{fig:fig5}(c)). In $J_{z}$, we notice one cycle of damped oscillation with a pair of peaks in opposite directions after the driving current has been switched off. However, we do not see two clear peaks in $P_{T}$.

From the spatial distribution of $V_{z}$ for $I_{0}$ = 0.36 kA, 1.98 kA and 3.60 kA at 13.9 $\mu$s, we find that $V_{z}$ increases with an increase of $I_{0}$ (see Fig.~\ref{fig:fig5}(a)). Just like $V_{z}$ and $P_{T}$ at ($x$,$y$) = [0,0] cm, we find that $V_{z}$ starts to increase from 7.6 $\mu$s. However, unlike $J_{z}$ and $P_T$, the first peak in $V_{z}$ is attained later than 13.9 $\mu$s for all values of $I_{0}$ (denoted by the first black dashed vertical line in Fig.~\ref{fig:fig5}(d)). Thus, there is a time lag for $V_{z}$ in comparison to $J_{z}$ and $P_{T}$. Furthermore, the peaks in $V_{z}$ are not attained at earlier times for higher $I_{0}$ than those for lower $I_{0}$. Following the initial time lag, $V_{z}$ reaches zero at a time later than 24 $\mu$s. Also, in contrast to $J_{z}$ and $P_{T}$, for higher values of $I_{0}$ (i.e., for $I_{0}$ = 2.52 kA, 3.06 kA and 3.60 kA), the magnitude of $V_{z}$ during the second stage of reconnection is larger than during the first stage with a $y$-directed current sheet (see Fig.~\ref{fig:fig5}(d)). However, in general, similar to $J_{z}$ and $P_{T}$, the magnitude of $V_{z}$ at ($x$,$y$) = [0,0] cm increases with $I_{0}$  in both the first and second stage (see Fig.~\ref{fig:fig5}(d)). After the driving current is switched off, $V_{z}$ undergoes a one cycle damped oscillation before vanishing at around 60 $\mu$s. In these late stages, while the profiles of $J_{z}$ and $P_{T}$ tend to overlap (see profiles in cyan shaded regions in Figs.~\ref{fig:fig5}(b)-(c)), the profiles of $V_{z}$ are more separate (see Fig.~\ref{fig:fig5}(d)). 
 
\section{ Discussion and Conclusion}
We have simulated oscillatory reconnection driven by alternating current-induced fast magnetoacoustic waves in laboratory plasma conditions. The mutually beneficial coexistence of reconnection and waves is important in the solar corona where the waves can drive reconnection and waves can be generated by reconnection \citep[e.g.,][]{Mondal2024b,Srivastava2025}. The present study indicates that such a symbiosis of waves and reconnection may be relevant and significant also at the laboratory scales. We find that oscillatory reconnection can also take place in the absence of Hall effects. However, the introduction of the Hall effect generates additional features such as in-plane Hall currents, quadrupolar magnetic fields and plasma flows in the out-of-plane direction. Furthermore, strong reconnection plasma outflows generates inverse Hall currents in the outflow regions which lead to more complex configurations of $B_{z}$. All of these features are present in a laboratory experiment that studies fast wave-driven formation of a current sheet by collapse of a magnetic null \citep[e.g.,][]{Frank2024}.

We measure the variation in time of  physical variables such as $J_{z}$, $P_{T}$ and $V_{z}$ at the magnetic null location for seven amplitudes of the driving current from 0.36 kA to 3.60 kA. We find that an increase in $I_{0}$ raises the values of $J_{z}$, $P_{T}$ and $V_{z}$ which possess nearly sinusoidal trends up to 44 $\mu$s. However, $V_{z}$ possesses a time-lag compared with the driving current, whereas $J_z$ and $P_T$ peak almost at the same time as the driving current or even slightly earlier. The higher mass of ions compared with electrons implies smaller accelerations for the ions \citep[e.g.,][]{Mandt1994,Huba1995,Rogers2001}, which produces the time-lag in $V_{z}$. Thus, the bulk plasma possesses a finite response-time to the Hall-modified magnetic stresses. 

Moreover, $J_{z}$ and $P_{T}$ attain higher peak values in the first stage of reconnection with a $y$-directed current sheet. In contrast, $V_{z}$ attains higher values in the second stage for higher values of $I_{0}$. Actually, in the initial stage, the collapse of the null is driven in the absence of prior Hall currents. Therefore, the entire wave energy is available to distort the null and thin the current sheet. As time advances, the presence of the $y$-directed current sheet breaks the initial symmetry, which in turn accelerates the decoupling of ions and electrons \citep[e.g.,][]{Mandt1994,Huba1995,Rogers2001}. Hence, the Hall effects gradually start to become more dominant and to modify the background conditions and introduce dispersive effects. Therefore, the $x$-directed current sheet tends to possess a lower current $J_{z}$, while $J_{z}$ and $P_{T}$ attain lower values in the second phase than in the initial phase. The ions are accelerated at a slower rate during first-stage of reconnection within the $y$-directed current sheet \citep[e.g.,][]{Mandt1994,Huba1995,Rogers2001}, but, during the second phase, $V_{z}$ attains higher peaks for higher $I_{0}$.

Oscillatory reconnection has been studied extensively under solar coronal conditions using resistive MHD simulations in 2D \citep[e.g.,][]{Craig1991, McLaughlin2009, McLaughlin2012,Karampelas2023} as well as in 3D \citep[e.g.,][]{Thurgood2017} in the absence of Hall physics. From these studies, it has been found that, if the null is perturbed by a single fast wave-like pulse, oscillatory reconnection can continue for multiple cycles, either due to the reflective nature of the boundaries \citep[e.g.,][]{Craig1991} or even  in the absence of such reflection \citep[e.g.,][]{McLaughlin2009}. Specifically, it has been shown that the period of oscillatory reconnection decreases with an increase in the amplitude of an aperiodic velocity driver \citep[e.g.,][]{McLaughlin2009}. Later, it was also reported that the period of oscillatory reconnection is inversely proportional to the background magnetic field and the square root of the initial plasma pressure \citep[e.g.,][]{Karampelas2023}. In present study, we have imposed fast waves continuously at the magnetic null from 4 $\mu$s to 44 $\mu$s. We find that the current sheet changes its orientation only once. So, the system behaves nearly in accordance with the driving current until it is switched off. After 44 $\mu$s, the system possesses a damped oscillatory behaviour in which the current sheet changes its orientation before reaching quasi-equilibrium. This oscillatory behaviour is due to the reflective nature of the boundaries \citep[e.g.,][]{Craig1991,Karampelas2022}.

Now, in solar coronal conditions, oscillatory reconnection has been found to continue for multiple cycles \citep[e.g.,][]{Craig1991, McLaughlin2009}. However, if the Lundquist number $S$ is less than $10^{4}$, the perturbation of a magnetic null will decay in less than one oscillation time \citep[e.g.,][]{Craig1991}. In general, it has been reported that, if the oscillatory reconnection is due to reflections from the boundaries, the time-scale for the oscillation is $\tau_{osc} = 2\ln S \times R/v_{A}$ s, where $R$ is the distance of the boundary from the position of the null and $v_{A}$ is the Alfv\'en speed at the boundary \citep[e.g.,][]{Craig1991}. The oscillation will decay over a time-scale given by $\tau_{decay} = (2/\pi^{2}) (\ln S)^{2} \times R/v_{A}$ s. In the present case, we have $R = 54~\mathrm{cm}$, and $v_{A} = 55 \times 10^{6}~\mathrm{cm.s^{-1}}$. The numerical magnetic diffusivity is $\eta_{numerical}=\delta x \times v_{A}$, where $\delta x$ is the smallest cell size. Since the reconnection dynamics is taking place symmetrically about the centre of the domain, the system size, say, $l$ can be considered to be the distance from the centre to the boundaries, namely, $l=R =54$ cm. Hence, the Lundquist number $S$ can be estimated to be 160. This makes $\tau_{osc}$ and $\tau_{decay}$ to be around 9.8 $\mu$s and 5 $\mu$s, respectively. From the estimated profiles of $J_{z}$, $P_{T}$ and $V_{z}$ in the absence of a driving current, we can see that the oscillation decays even before completing one oscillation. Thus, the late-stage behaviour in our simulation agrees qualitatively with what is expected for a plasma system with low Lundquist number. In future, it would be interesting to study the effects of changes in the initial background plasma density, temperature, magnetic field on the dynamics, as well as the effect of an increase in spatial resolution with an accompanying increase in effective Lundquist number. Carrying out a laboratory experiment on this scenario, as well as a fully kinetic simulation, would be invaluable.

\section*{Acknowledgments}
The authors are thankful to the anonymous reviewers for their valuable constructive remarks which have been very helpful in improving the manuscript. The authors appreciate the user-friendly, flexible framework of open source MPI-AMRVAC 3.0 which is helpful to perform this simulation in laboratory scales using Hall-MHD formalism. S.M. acknowledges the financial support provided by the Prime Minister's Research Fellowship (PMRF) of India. A.K.S would like to acknowledge the ISRO grant(DS/2B-13012(2)/26/2022-Sec.2) for the support of his scientific research.



\bibliographystyle{aasjournal}
\bibliography{Osc_Rec}



\end{document}